# Now You See It, Now You Don't: Adversarial Vulnerabilities in Computational Pathology

Alex Foote, Amina Asif*, [1], Ayesha Azam, Tim Marshall-Cox, Nasir Rajpoot and Fayyaz Minhas

Department of Computer Science, University of Warwick, Coventry CV4 7AL, UK
* amina.asif@warwick.ac.uk

**Abstract.** Deep learning models are routinely employed in computational pathology (CPath) for solving problems of diagnostic and prognostic significance. Typically, the generalization performance of CPath models is analyzed using evaluation protocols such as cross-validation and testing on multi-centric cohorts. However, to ensure that such CPath solutions are robust and safe for use in a clinical setting, a critical analysis of their predictive performance and vulnerability to adversarial attacks is required, which is the focus of this paper. Specifically, we show that a highly accurate model for classification of tumour patches in pathology images (AUC > 0.95) can easily be attacked with minimal perturbations which are imperceptible to lay humans and trained pathologists alike. Our analytical results show that it is possible to generate single-instance white-box attacks on specific input images with high success rate and low perturbation energy. Furthermore, we have also generated a single universal perturbation matrix using the training dataset only which, when added to unseen test images, results in forcing the trained neural network to flip its prediction labels with high confidence at a success rate of > 84%. We systematically analyze the relationship between perturbation energy of an adversarial attack, its impact on morphological constructs of clinical significance, their perceptibility by a trained pathologist and saliency maps obtained using deep learning models. Based on our analysis, we strongly recommend that computational pathology models be critically analyzed using the proposed adversarial validation strategy prior to clinical adoption.

## 1 Introduction

Over recent years, deep learning has been used to model a number of problems in computational pathology (CPath) like cancer detection and classification, cell detection and classification, predictive and prognostic analyses using whole slide images etc. [1–5]. In CPath and related fields, the ultimate goal is to develop systems that can be deployed in clinical settings to assist human experts [6]. As these systems are intended to be involved in making decisions that could have an impact on patients' health, rigorous performance evaluation is needed before their deployment [7]. Typically, machine learning (ML) models are evaluated over accuracy-based metrics using cross-validation and/or independent set testing [8]. A model is deemed to be suitable for real world use if it shows decent accuracy during such performance evaluation.

---

[1] The authors wish it to be known that, in their opinion, the first two authors should be regarded as joint first authors.

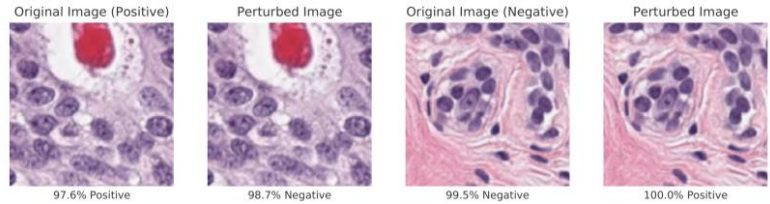

**Fig. 1.** Two examples of correctly classified images and their misclassified perturbed versions

Traits like automated feature extraction and the ability to learn complex problems make deep learning a favored choice among data scientists working on CPath problems [9]. However, there is significant evidence in the literature that demonstrates that deep models are vulnerable to adversarial attacks, i.e., perturbations that might be visually imperceptible but cause a model to change its classification of an image [10–14]. Robustness of a model can be quantified by analyzing the perturbations required to successfully attack it [15]. A model that exhibits susceptibility to low energy perturbations might hint at overfitting during training [16].

Whole slide images in CPath are inherently susceptible to small changes due to factors like compression, rescanning, differences in scanning equipment, staining protocols and environment, etc. [17, 18]. Even when there is no malicious intent to deliberately fool a system, a model may encounter examples that are visually similar to the samples it was tested on but contain slight variations in intensity values. Since creating test sets that capture all possible variations is not practically possible, we need alternate ways to test and quantify robustness of CPath models to ensure their suitability for clinical use.

In this paper, we propose performing vulnerability analysis of trained CPath models using adversarial attacks in addition to conventional accuracy-based evaluation. To demonstrate that using only accuracy-based metrics might not suffice in capturing the overall robustness of a model, we developed a ResNet-18 based tumour classifier and performed cross-validation analysis [19]. We then generated and analyzed the effects of single-instance and universal adversarial attacks over the model. Two examples of successful single-instance attacks are shown in **Fig. 1**. We also evaluated their impact on a trained pathologist to determine the susceptibility of human experts to adversarial attacks in comparison to a deep learning model. To the best of our knowledge, this is the first systematic evaluation of robustness analysis using adversarial attacks in CPath models.

## 2   Methods

In this study, we propose the use of adversarial attacks to analyze robustness of deep learning models for CPath in addition to conventional accuracy-based performance assessment. We demonstrate that despite showing high accuracy, a model may be susceptible to adversarial attacks, hence raising questions over its suitability for deployment in clinical settings. To illustrate this phenomenon, we first train a Convolutional Neural Network (CNN) for detecting tumour cells within a given patch of a hemotoxylin and eosin (H&E) stained whole slide image. After evaluating the model's classification performance through traditional performance metrics, we generate and analyze the effects of adversarial attacks over it. The following sections present details of model training, adversarial attacks and experimental setup employed in this study.

### 2.1 Tumour Cell Classification

We developed a classification model for tumour cell detection as a case study for evaluating the robustness of CPath models. Classification of whole slide image patches containing tumour cells is an important problem in CPath with important applications in cancer diagnostics [20]. To develop our base classification model for this study, we used a subset of the PanNuke [21] dataset which contains patches of tissue images from 19 different sites annotated for 5 different cell classes (Epithelial, Neoplastic, Necrotic, Inflammatory, and Connective) with a total of >200,000 nuclear level annotations. Each 224×224 pixel image patches in the dataset is obtained from whole slide images scanned at a resolution of 0.25 microns per pixel. To limit the impact of cross-tissue morphological variations, we used only the breast tissue subset of the dataset and labeled the images based on their tumour cell composition. A patch was labeled positive if it contained 5 or more tumour cells and negative if there were no tumour cells in it. This resulted in a dataset of 1,157 positive and 878 negative class patches.

In line with existing transfer-learning strategies employed in a variety of CPath problems [22], we used a ResNet-18 [19] pretrained on ImageNet [23] with two neurons with softmax activation in the final layer for modeling this binary classification problem [24]. We implemented the model in PyTorch [25] and trained using the AdaDelta [26] optimizer with a learning rate of 0.01 for 25 epochs. For performance assessment, we used 5 runs of 3-fold stratified cross-validation and calculated the mean accuracy and AUC-ROC across the runs. To visualize important regions in images for classification, we computed their gradient-based saliency maps [27, 28] with the trained model. We also analyzed the features at the penultimate layer of the network through Uniform Manifold Approximation and Projection (UMAP) [29] and the loss landscape of the neural network through Principal Component Analysis (PCA) of the weight parameters of the neural network model to ensure convergence to a suitable minima during training.

### 2.2 Adversarial Attacks on Computational Pathology Models

**Single-instance Attacks**

In this section, we describe the algorithm used for generating single-instance attacks on the tumour detection network. For a given input image $x$, the aim of a single-instance adversarial attack is to find a perturbation $\delta$ so that the perturbed image $x' = x + \delta$ is very similar to the original image $x$ but is classified to a different class by the trained neural network $f(\cdot; \theta)$ with high confidence. This corresponds to finding the perturbation $\delta$ that maximises the value of the loss function $l(\cdot)$ between the true label $y$ and the model's prediction on $x'$:

$$\max_{\delta} l(f(x + \delta; \theta), y). \tag{1}$$

To find a good approximation of this problem, we used Projected Gradient Descent (PGD). Mądry et al. [10] found that PGD was successful at finding a strong perturbation suitable for adversarial training, which makes it appropriate for assessing the robustness of a model to a worst-case perturbation. At each step, PGD takes the gradient of the loss function with respect to $\delta$ and updates the values $\delta$ in the positive gradient direction, to maximize the loss $l$. This can be expressed formally as: $\delta_{i+1} = \delta_i + \gamma \nabla_\delta l(f(x + \delta_i; \theta), y)$. Here, $\nabla_\delta l$ is the gradient of the loss function with respect to the perturbation $\delta_i$ and $\gamma$ refers to the learning rate. This is repeated until the label has been flipped with sufficient confidence, or after a given number of steps, which implicitly constrains the magnitude of the perturbation.

**Constraints on Perturbation Energy**

To ensure that the algorithm finds an adversarial input that is sufficiently similar to the original input, a constraint over the magnitude of the perturbation can be applied [30]: $\|\delta\|_p < C$. This specifies that the $L_p$ norm of the perturbation $\delta$ must be less than a chosen value $C \geq 0$. The constraint is applied after each gradient step. We evaluated our model under the four most common constraints, the $L_0, L_1, L_2$, and $L_\infty$ norms. We additionally performed evaluation with an unconstrained attack. For this problem, the $L_0$ constraint over $\delta$ corresponds to limiting the proportion of pixels that are changed as the result of the perturbation. We applied this constraint by finding the top $c$ portion of the pixels with the largest gradient values and updating only these pixels. Similarly, $\|\delta\|_1$ measures the Mean Absolute Difference (MAD) between $x'$ and $x$. Given a maximum allowable MAD, the perturbation was rescaled such that the MAD is less than or equal to the constraint. The $L_2$ constraint on the perturbation is expressed as the Root Mean Squared Difference and applied analogously. The $L_\infty$ constraint captures the maximum difference between any pair of pixels in $x'$ and $x$. Given a maximum allowable difference, the perturbation was clamped to this value in the positive and negative directions.

**Universal Attacks**

In addition to the single-instance attacks produced by finding an optimal perturbation for each image separately, we generated a single perturbation that could be used to successfully attack multiple images [31]. For this purpose, we utilized the training dataset to generate a single perturbation matrix that can be used to produced a successful attack when added to unseen or novel images. Given a labeled training dataset of $n$ examples $\{(x_i, y_i) | i = 1 \ldots n\}$, we formulate the optimization problem of finding an optimal universal perturbation matrix $\delta^*$ as:

$$\delta^* = \underset{\delta}{argmax} \sum_{i=1}^{n} l(f(x_i + \delta; \theta), y_i). \tag{2}$$

As opposed to single-instance attacks, the above formulation results in one attack matrix by maximizing the sum of the losses over $n$ training examples. We approximate a solution to this problem using Stochastic Gradient Descent (SGD), updating the perturbation after each training example rather than over the entire training set. The resulting universal matrix is then added to unseen test images to evaluate the success rate of these attacks.

### 2.3 Perceptibility Analysis of Adversarial Attacks by Pathologist

To assess the visual impact of the adversarial perturbations on the images, we evaluated the effect of the attacks on the decisions of a human pathologist. This aimed to find if successful adversarial attacks introduce perceptible changes in the images that might cause a human expert to change their decision about the associated labels, similarly to the model. As detecting and counting tumour cells in an image is a time-consuming task that requires long periods of visual attention, we selected a subset of 20 positively labeled examples from the test set that were originally classified correctly by the classification model and were successfully attacked, i.e., their perturbed versions were classified incorrectly by the model. To minimize the effect of intra-observer variation in labeling we created two copies of each original-perturbed image pair and asked a pathologist to label all the 40 images using the same labeling criterion used in the construction of this dataset. Furthermore, the pathologist was kept blind to the original labels in the dataset. We quantified the impact of adversarial attacks on the human expert by comparing the number of True Positives (TPs) for both the original and perturbed images.

## 3 Experiments and Results

### 3.1 Standard Model Evaluation

The cross-validation results of the machine learning model used for classifying patches into tumour and non-tumour patches across five runs are given in **Table 1** a). This network gives area under the Receiver Operating Characteristic curve (AUC-ROC) of > 96% with a mean accuracy of 89%. These results seem to indicate that the proposed model generalizes well to unseen test cases. We also analyzed the saliency maps generated by the machine learning model (see **Fig. 3**), which show the base neural network model is picking up tumour cells within test image patches. **Fig. 2** shows the two-dimensional UMAP plot of the features corresponding to the second last layer of the base network along with the loss landscape of the neural network with its training trajectory overlayed on it. This analysis clearly shows that the neural network has learned features that discriminate between tumour and non-tumour patches over an unseen test set and that the neural network is in a stable optimization region with low loss. Such analyses would normally be considered sufficient for determining that the CPath model is suitable for the task.

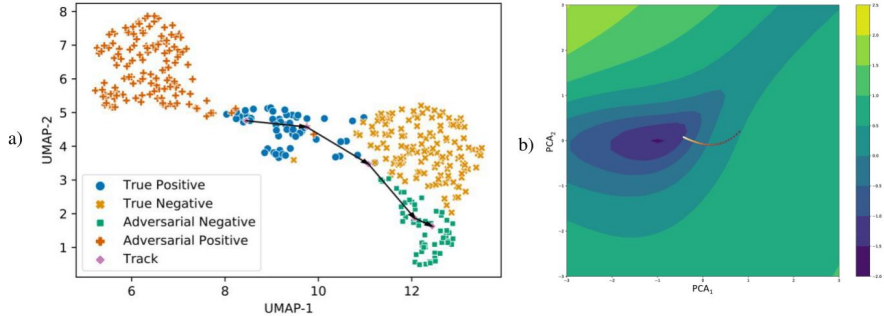

**Fig. 2.** (a) UMAP of penultimate layer features: True positives (negatives) are unperturbed inputs classified as positive (negative) by the model, and adversarial positives (negatives) are perturbed inputs classified as positive (negatives). The "track" shows the optimization path of an image during the attack. (b) PCA-based loss landscape of base model and its training trajectory.

**Table 1.** a) Model accuracy and AUC-ROC averaged over three folds and five runs, b) Pathologist evaluation results – the scores are the True Positives, with a maximum of 20.

|  | Mean | Standard Deviation |
|---|---|---|
| Accuracy | 0.89 | 0.014 |
| AUC-ROC | 0.96 | 0.004 |

a)

|  | Original | Perturbed |
|---|---|---|
| Set 1 | 17 | 18 |
| Set 2 | 18 | 18 |

b)

### 3.2 Evaluation with Single-instance Attacks

**Fig. 4** shows the success rate of the attacks when varying the $L_p$ norm constraints. Even under strict constraints the attacks are successful at flipping the classified label 100% of the time. This shows that the model is highly vulnerable to adversarial attacks even at low perturbation energies, where the perturbations are imperceptible to humans. In fact, under the $L_\infty$ constraint, the attacks are all successful even with a maximum allowed pixel perturbation of 0.5, the minimum perturbation required to change a pixel value. Additionally, under the $L_0$ constraint, the model is consistently deceived when only 20% of the pixels are changed, showing that even when most pixels are unchanged a perturbation can cause misclassification.

### 3.3 Evaluation with Universal Attacks

In addition to single-instance attacks, we also generated a single universal perturbation to attack multiple images. For simplicity, we generated the perturbation over the positive training examples only and evaluated the perturbation on the positive test examples. The perturbation was successfully able to attack 619 out of 702 (88.2%) images in the training set. When it was used to attack previously correctly classified test images of the positive class, we observed success in 319 out of 376 (84.8%) cases. As shown in **Fig. 5**, no visible structural changes or artifacts were observed in the perturbed images.

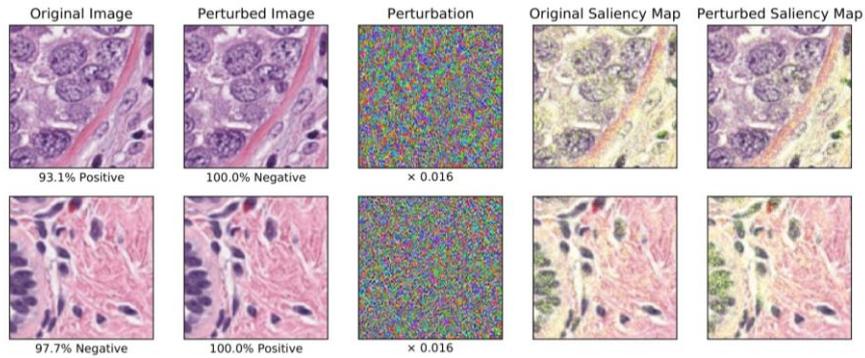

**Fig. 3.** Examples of an original image, the image with a single-instance perturbation applied, a visualization of the perturbation, and saliency maps before and after the perturbation

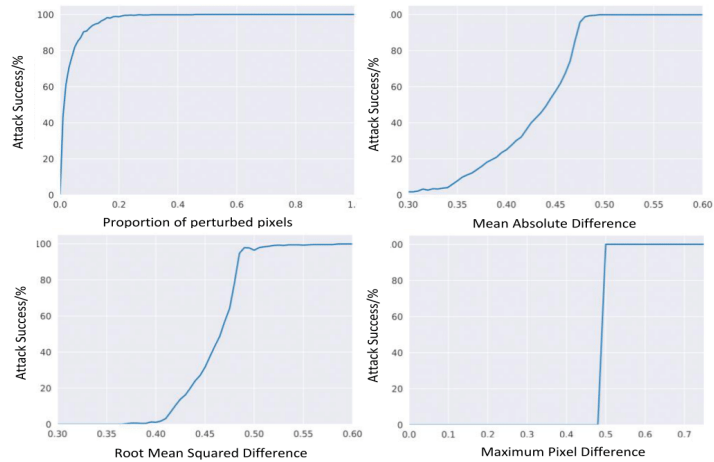

**Fig. 4.** Attack success rate under (from left to right) $L_0, L_1, L_2,$ and $L_\infty$ constraints

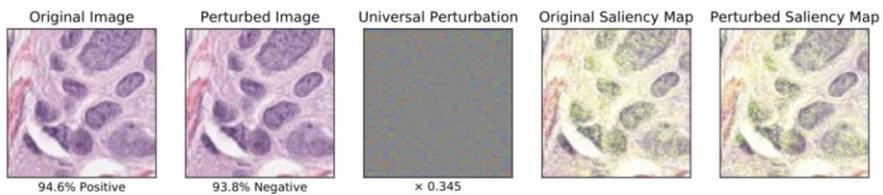

**Fig. 5.** An example of an original and perturbed image, with a visualization of the perturbation and the original and perturbed saliency maps

### 3.4 Interpreting Adversarial Attacks

**Fig. 2** a) shows the UMAP projection of the feature representations for images before and after single-instance attacks. Notably, the adversarial negatives are closer in the space to the true negatives than to the true positives, and vice versa. The path of an image during the attack process is shown, with a point plotted after each optimization step. The image begins in the original true positive class, and moves through the space towards the negative points as the perturbation is applied, eventually settling in the adversarial negative group. These observations provide further evidence that the adversarial perturbations genuinely fool the model into misclassifying the images, rather than simply applying sufficient noise to disrupt the classification.

**Fig. 3** and **Fig. 5** show comparisons between the saliency maps of unperturbed images and their perturbed partners. The saliency maps extracted from the model on unperturbed images suggest that it is learning desirable features that place emphasis on parts of the images that contain useful information, such as tumour cells. However, the maps extracted for perturbed inputs exhibit clear differences, showing that the adversarial attack caused a change in the pixels to which the model attends. It appears that when the model is fooled into misclassifying a positive example as negative, the attention of the model is shifted away from the tumour cells. In contrast, when the model is fooled into misclassifying a negative image as positive, attention is shifted towards the non-tumour cells, suggesting the model now perceives these as tumourous.

This analysis shows that convincing saliency maps do not provide sufficient evidence to conclude that the model is learning robust features that predict the desired outcome and is instead also using non-robust features that can be easily perturbed. This corresponds to the work of [32, 33], which suggested that non-robust models are able to use high-frequency features to improve their prediction performance, at the cost of leaving them vulnerable to attacks that can easily modify such features to cause misclassification.

## 4    Conclusions and Recommendations

There is substantial evidence to show that ML models can exploit unexpected properties of a task to achieve high accuracy, including in the context of medical imaging [34]. Whilst adversarial robustness is not sufficient to ensure that the model is learning desirable features, a lack of robustness clearly demonstrates that the model fails in non-human ways, providing a warning flag to encourage further analysis. Our results show that adversarial attacks on CPath models can reveal such vulnerabilities and should be used in conjunction with other performance evaluation protocols in the validation of such solutions prior to clinical adoption. We show that minimal perturbations can fool an otherwise accurate predictive model for tumour patch classification, and possibly for other related problems as well. Based on our analyses, we suggest that developing CPath solutions using human interpretable features can possibly improve both interpretability of the model and its robustness to adversarial or natural perturbations due to factors such as image compression, scanner variation, staining differences, etc. Furthermore an adversarial training paradigm [35] can also help to ensure that the model's decision is based on meaningful differences amongst classes rather than opaque and potentially spurious reasons.